\documentclass[aps,pra,twocolumn,showpacs,superscriptaddress]{revtex4}
\usepackage{graphicx}
\usepackage{dcolumn}
\usepackage{color}
\usepackage{bm}
\usepackage{amssymb}
\usepackage{amsmath}
\begin{document}
\title{A study on evolution of a cold atom cloud in a time dependent radio frequency dressed potential}

\author{A. Chakraborty}
\email[E-mail: ]{carijit@rrcat.gov.in}
\author{S. R. Mishra}
\affiliation{Raja Ramanna Centre for Advanced Technology, Indore-452013, India.}
\affiliation{Homi Bhabha National Institute, Mumbai-400094, India.}
\author{S. P. Ram}
\author{S. K. Tiwari}
\affiliation{Raja Ramanna Centre for Advanced Technology, Indore-452013, India.}
\author{H. S. Rawat}
\affiliation{Raja Ramanna Centre for Advanced Technology, Indore-452013, India.}
\affiliation{Homi Bhabha National Institute, Mumbai-400094, India.}
\begin{abstract}
Using a Direct Simulation Monte Carlo technique, we have studied the time evolution of a cold atom cloud interacting with a time dependent radio frequency (rf) dressed state potential. Exposure of a cloud of $^{87}Rb$ atoms, trapped in a quadrupole magnetic trap, to a time dependent rf-field with increasing amplitude and decreasing frequency, shows a variation in the number of trapped atoms and the overall shape of the atom cloud. It is shown by simulations that, initially at lower rf-field strength, the rf-field results in ejection of atoms from the trap, leading to evaporative cooling of the atom cloud. Later, at higher rf-field strength, the atoms undergo the non-adiabatic Landau-Zener (LZ) transitions, which leads to their trapping in an rf-dressed state potential of toroidal shape. The results of simulations explain the experimentally observed results. The simulations can be useful to predict the atom cloud dynamics and trapping geometries with other forms of the potential.
\end{abstract}

\maketitle

\section{Introduction}
The radio frequency dressed potentials (rf-dressed potentials), initially proposed by Zobay and Garraway \cite{Zobay:2001}, are considered as convenient tool to trap and manipulate ultra-cold atoms in exotic forms \cite{Petrov:2004,Lesanovsky:2006:73,Gildemeister:2010}. Non-trivial trapping geometries like ring trap, double well and shell traps have already been demonstrated using potentials generated from various combinations of static magnetic field, radio frequency (rf) field and far detuned laser beams \cite{Folman:2002,Colombe:2004,Morizot:2006,Morizot:2007,Lesanovsky:2007,Heathcote:2008,Morizot:2008,Sherlock:2011,Merloti:R:2013, Chakraborty:2016}. These trapping geometries provide opportunity to explore basic physics of fermions and bosons in low dimensions, matter wave interferometry \cite{ Berman:book, Andrews:1997}, etc. The ring shaped atom traps, besides being useful for fundamental studies, are also promising for the development of atom gyroscope for precision rotation sensors for navigation \cite{Segal:thesis,Canuel:2006,LeCoq:2006}.

The rf-dressed potentials, also known as ``adiabatic potentials'', are energy eigenvalues of an interaction Hamiltonian for an atom interacting with an rf-field in presence of a static magnetic field \cite{Zobay:2001}. Dependence of these potentials on the atom-field coupling strength and frequency detuning of the rf-field with the Zeeman sub-levels transition frequency, make them inherently position dependent and capable of providing interesting atom-trapping geometries \citep{Schumm:2005, Hofferberth:2006,Lesanovsky:2006:73,Chakraborty:2014}. The rf-dressed potentials also promise a great flexibility of tailoring them by changing the rf-field strength, polarization, frequency and phase \cite{Chakraborty:2014,Chakraborty:2016}. By varying the rf-field parameters, the resultant rf-dressed potential can be made time dependent, which may be exploited for multiple purposes, such as initially for evaporative cooling of atom cloud and then for trapping in a designed geometry. Such an example of atom trapping in a toroidal geometry has been demonstrated recently by our group by applying an rf-field of varying amplitude and frequency on a cloud of $^{87}Rb$ atoms trapped in a quadrupole magnetic trap \cite{Chakraborty:2016}. 

In the earlier work \cite{Chakraborty:2016}, it was demonstrated that a quadrupole magnetic trap was converted into a toroidal shaped atom-trap, when the frequency and amplitude of the applied rf-field were ramped (varied with time) in the chosen way. In that work, the atom cloud was characterized and studied after the end of the ramp of the rf-field - when rf-field was left ON at a fixed final frequency and amplitude. A natural curiosity, however, related to these experiments, remains to understand the various processes involved and their effect on the evolution of the characteristics of the atom cloud during the ramp of the rf-field. In the present work, we have made an attempt to understand the time evolution of the atom cloud, in terms of variation in its shape and number of atoms, during the rf-field ramp (i.e. during the transformation of the quadrupole trap into an rf-dressed toroidal trap). We have performed the simulations considering the atom cloud evolving in a time dependent potential formed due to ramp of rf-field in presence of a static quadrupole magnetic field. The simulations results have been compared with the experimental observations. 

In the simulations, we have used the Direct Simulation Monte Carlo (DSMC) technique which is based on an extension of the original proposal of Bird \citep{Bird:1994}. Different variants of DSMC method have already been used to understand the evaporative cooling \cite{Wu:1996,Wu:1997} and expansion cooling \cite{Wu:1998} of atoms. The results of simulations show that, at lower rf-field strength and higher frequency (at the beginning of ramp of the rf-field), the rf-field results in ejection of atoms from the trap, leading to evaporative cooling of the atom cloud. Later on, at higher rf-field strength and lower frequency, the atoms can undergo the Landau-Zener (LZ) transition which leads to their trapping in an rf-dressed potential of toroidal shape. The results of these simulations explain well our experimental observations related to temporal variation in the number of atoms and shape of the atom cloud in the trap. Similar simulations can be performed to predict the dynamics of atom cloud with more sophisticated evolution of the potential which can help in exploring more efficient cooling and trapping schemes.

The DSMC method does not require the ``sufficient ergodicity" approximation and thus can simulate atomic ensembles far from their equilibrium condition. This makes DSMC technique more suitable to simulate our experimental conditions where a fast ramp of the rf-field parameters (amplitude and frequency) is considered. The ergodicity assumption also fails if the mixing time of the atomic samples is longer than the elastic collision time \cite{Wade:2011}. For other situations, both classical and quantum kinetic theories of truncated Boltzmann distribution of the phase space provide the alternative approaches to the evaporative cooling mechanism \cite{Luiten:1996,Yamashita:1999}. But these treatments are valid with the assumption of the ``sufficient ergodicity", which means that the phase space density is only a function of energy.

The article is organised as follows. The section \ref{Theory and simulations} of this article presents the theoretical background related to the study along with the method of numerical simulations. The experimental procedure, observations, results of numerical simulations and the comparisons between experiments and simulations have been presented in \ref{Results}. The conclusion of our work is discussed in section \ref{Conclusion}.

\section{Theory and simulations}\label{Theory and simulations}
\subsection{Trapping in rf-dressed potential}

The hyperfine state $|F=2\rangle$ of $^{87}Rb$ is five fold degenerate and splits into Zeeman sub-levels in presence of a static magnetic field. We consider the atoms are prepared in the $|F=2,m_F=2\rangle$ state to trap them in a quadrupole magnetic trap. It is assumed that the trapping magnetic field is not very strong and the energy gap among Zeeman sub-levels varies linearly with the field strength. Hence, an applied rf-field of appropriate frequency can be simultaneously resonant to all the transitions involving adjacent sub-levels. The rf-field is considered to be ramped to have time varying amplitude and frequency, as discussed specifically later, in a manner as reported in our recent experimental work \cite{Chakraborty:2016}. When the rf-field is sufficiently strong (\textit{i.e.} Rabi frequency is high), the atom-field interaction is described better in the ``dressed state" \cite{Zobay:2004} picture involving the diagonalization of both the static and coupling Hamiltonian. In our simulations, we first calculate the rf-dressed potential (using a formalism based on rotating wave approximation (RWA)) at any instant of time during the ramp of the rf-field. The collision between atoms is formulated in the degenerate internal state (DIS) approximation which neglects the Rabi oscillations during the collisions, as collision between two cold  $^{87}Rb$ atoms occurs locally in a sub-nanosecond time regime \cite{Tscherbul:2010,Hofferberth:2006}. This approximation makes it possible to express the scattering amplitudes between pairs of dressed atoms in terms of bare state scattering amplitudes discussed in detail in an earlier work \cite{Suominen:1998}. The process of rf-evaporation as well as trapping of the atoms in the rf-dressed potential has been studied numerically with the help of a DSMC algorithm, which simulates the motion of an atom in the time varying rf-dressed potential. In the simulations, the Landau-Zener (LZ) transition probability is calculated to ascertain if the atom is going to be trapped in the rf-dressed potential or not. Depending upon this probability and the coordinates of the atom evolved in the time dependent potential, the final atom cloud trapped in the rf-dressed potential is constructed.  

We consider the static quadrupole magnetic trap with magnetic field distribution near the trap centre given by $\textbf{B}^\textbf{S}\textbf{(r)}=B_q\left(x,y,-2z\right)$, where $B_q$ is the radial field gradient. In this trap, the transition frequency between adjacent Zeeman sub-levels is given as, 
\begin{equation}
\omega_0=\frac{g_F\mu_BB_q}{\hbar}\sqrt{x^2+y^2+4z^2},
\end{equation}
where $g_F$ is the Lande's g-factor, $\mu_B$ is the Bohr magneton and $\hbar$ is the reduced Planck's constant. We consider an atom interacting with a static field $\textbf{B}^{\textbf{S}}(\textbf{r})$ and an rf-field of frequency $\omega$ with field vector components $\textbf{B}^\textbf{{rf}}(t)=\left(B^{rf}_{\perp 1}(t),B^{rf}_{\perp 2}(t),B^{rf}_{||}(t)\right)$, where $B^{rf}_{\perp 1}(t)$, $B^{rf}_{\perp 2}(t)$ and $B^{rf}_{||}(t)$ are the perpendicular and parallel components of the rf-field respectively with respect to the local static field vector $\textbf{B}^{\textbf{S}}(\textbf{r})$. The interaction Hamiltonian for the atom-field interaction is given as \cite{Hofferberth:2007} ,
\begin{equation}\label{eq:H}
H(t)=\frac{g_F\mu_B}{\hbar}\left[\sum_{i=1,2}F_iB^{rf}_{\bot i}(t)+F_3\left(|\textbf{B}^\textbf{S}|+B^{rf}_{||}(t)\right)\right].
\end{equation}
where $F_i$ ($i=1,2$) are the hyperfine operators along the $B^{rf}_{\perp 1}(t)$ and $B^{rf}_{\perp 2}(t)$ components and $F_3$ is the hyperfine operator along $B^{rf}_{||}(t)$.

\begin{figure}[t]
\includegraphics[width=8.2cm]{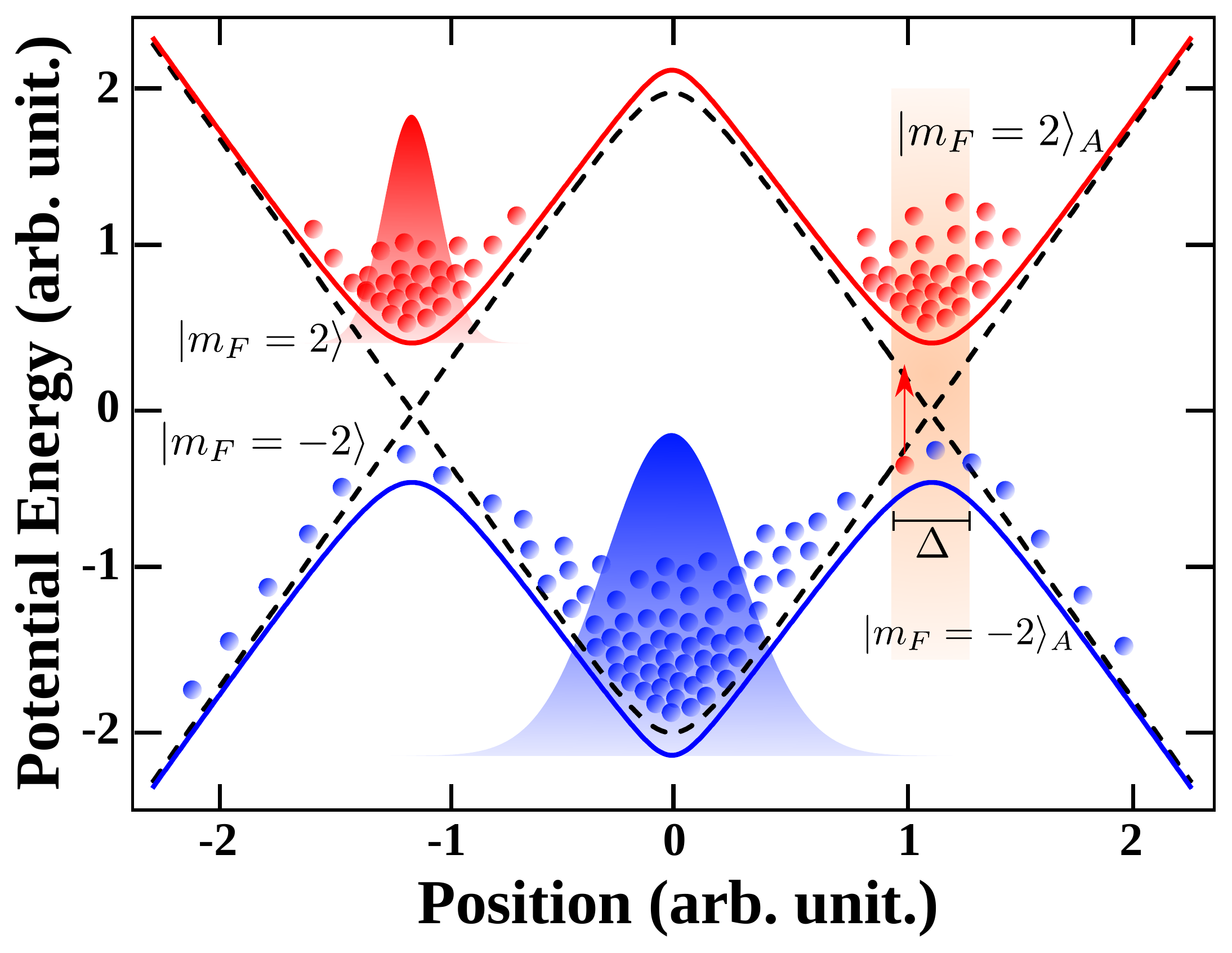}
\caption{\label{fig:field_line}(Color online) Schematic of the variation in potential energy of an atom with position in a quadrupole trap in presence of an rf-field. The continuous curves show the adiabatic rf-dressed potentials (denoted with the subscript A) including the coupling of rf-field with atom. The dashed lines show the atom-field energy without coupling of rf-field with atom. The red dots correspond to the atoms in $|m_F=2\rangle_A$ state while the blue dots correspond to atoms in $|m_F=-2\rangle_A$ state. The width $\Delta$ represents the interaction region where LZ transition probability is high. For brevity, the potentials for only two states are shown here.}
\end{figure}

If rf-field is taken to be of the form $\textbf{B}^\textbf{{rf}}(t)=\left[B^{rf}_{||}cos(\omega t), B^{rf}_{\perp 1}cos(\omega t-\gamma_1), B^{rf}_{\perp 2}cos(\omega t-\gamma_2)\right]$, then, using a semi classical treatment under rotating wave approximation (RWA), the interaction Hamiltonian of eq. (\ref{eq:H}) can be diagonalised to lead the energy eigen values (\textit{i.e.} rf-dressed potentials) given as,
\begin{equation}\label{eq:pot_compact}
V(\textbf{r})=m_F\hbar\sqrt{\delta^2+|\Omega|^2},
\end{equation}
where
\begin{equation}\label{eq:delta}
\delta(r)=\omega-\omega_0,
\end{equation}
\begin{equation}\label{eq:omega}
|\Omega|^2=\left(\frac{g_F\mu_B}{2\hbar}\right)^2{\left[2B^{rf}_{\bot 1} B^{rf}_{\bot 2}\sin\gamma+\sum_{i=1,2}(B^{rf}_{\bot i})^2\right]},
\end{equation}
and $m_F=$ -F to F for a given hyperfine state $|F\rangle$. Here $\delta$ is the detuning, $\Omega$ is the coupling strength (\textit{i.e.} Rabi frequency) for the adjacent Zeeman sub-levels and $\gamma=|\gamma_1-\gamma_2|$. Knowing $\delta$ and $\Omega$, the potential $V(r)$ can be evaluated using Eqs.(\ref{eq:pot_compact})-(\ref{eq:omega}). 
The eigen states (\textit{i.e.} rf-dressed states), corresponding to these eigen values (\textit{i.e.} rf-dressed potentials $V(r)$) can be written in compact form as,
\begin{equation}\label{eq:states}
|m_F=j\rangle_A = \sum_{k=-F}^{F}C_k^j|m_F=k\rangle,\ for\ j=-F,...,F
\end{equation}
where $|m_F=k\rangle$ are the bare sates corresponding to Zeeman sub-levels of hyperfine level $F$, $|m_F=j\rangle_A$ denotes the dressed states and coefficients $C_k^j$ are the matrix elements of a unitary rotation matrix which transforms the bare states to the dressed states. The values of few $C_k^j$ are given in the Appendix in terms of the detuning and Rabi frequency. A schematic of the variation in rf-dressed potential energy of an atom with position is shown in Fig. \ref{fig:field_line} for two dressed states $|m_F=-2\rangle_A$ and $|m_F=2\rangle_A$. The region with width $\Delta=\frac{\hbar\Omega}{\mu_Bg_FB_q}$ represents the interaction region where LZ transition probability is high. This region where potential energy curves for two dressed states come close to each other but do not cross each other, unlike the bare state energy curves, is known as the ``avoided crossing''.

Further, eq. (\ref{eq:H}) can be extended to include the kinetic energy of the atoms. The inclusion of atomic velocity introduces non-adiabaticity in the system which allows the transition between different dressed states near the avoided crossing. These non-adiabatic transitions have been addressed first by Landau and Zener for a two level system and later by \citet{Vitanov:1997} for a multi-level atomic system. The non-adiabatic Landau-Zener transition probability within Zeeman sub-levels of a hyperfine state $F$ can be written as \cite{Morizot:2006},
\begin{equation}\label{eq:PLZ}
P_{LZ}=(1-exp(-\pi\Lambda))^{2F}
\end{equation}
where $\Lambda=\frac{\hbar\Omega^2}{2g_F\mu_BB_qv}$ and $v$ is the velocity of the atom with which it traverses the avoided crossing. 

Irradiation of the rf-field transforms the initial trapping state $|m_F=2\rangle$ in quadrupole magnetic trap to $|m_F=-2\rangle_A$  of the dressed quadrupole trap in which atom experiences a finite potential depth due to the presence of the avoided crossings. The atoms with sufficient velocity can climb this potential and reach the avoided crossing region, where they either make a non-adiabatic LZ transition to the other dressed state $|m_F=2\rangle_A$ or adiabatically follow the $|m_F=-2\rangle_A$ state leading their ejection from the trap. In the low coupling regime, $\Omega\rightarrow 0$, the low LZ transition probability $P_{LZ}$ favours the removal of the atoms at the avoided crossing. At a higher coupling strength, a larger $P_{LZ}$ favours more number of atoms to make the non-adiabatic LZ transition to the other dressed state and get trapped in the trapping potential offered by that particular state. Hence by tailoring the coupling strength, we can either conduct an evaporative cooling of the atom cloud trapped in a quadrupole trap or transfer them to another potential landscape of rf-dressed state (an ellipsoid in case of $|m_F=2\rangle_A$). 

A physical equivalence in the bare state and dressed state pictures is as following. In the bare state picture of evaporative cooling, atom in $|m_F=2\rangle$ state, while being at the resonance to the rf field, makes a transition to the $|m_F=-2\rangle$ state and flies out of the trap. Whereas in dressed state formalism, evaporation of atoms is accomplished by only following the $|m_F=-2\rangle_A$ state, without considering any transition. In this picture, the non-adiabatic transition of an atom from the $|m_F=-2\rangle_A$ state to state $|m_F=2\rangle_A$ leads to trapping of the atom in a different potential landscape corresponding to state $|m_F=2\rangle_A$. 

The particular state $|m_F=2\rangle_A$ has potential minima on the surface of an ellipsoid co-centred with the quadrupole trap. With a suitable choice of strength of static magnetic and rf- fields, and the high confinement along the quadrupole trap axis (\textit{i.e.} z-axis), the trapping potential can form a toroidal trapping geometry. The atoms with low velocity and at high coupling strength have high probability to undergo LZ transition to the state $|m_F=2\rangle_A$ at the avoided crossing and get trapped in the minimum of this toroidal potential to form the ring shaped cloud. Here it can be noted, an atom interacting with the rf-field is likely to evolve in all the dressed states ($|m_F=0, \pm 1, \pm 2\rangle_A$) given by the eq. (\ref{eq:states}). But the mechanism of evaporation and trapping has been described by considering only the initial and final states ($|m_F=\pm 2\rangle_A$) due to the fast evolution of the spin states into these two states.

\subsection{Numerical simulation}\label{simulation}
The process of rf-evaporation as well as trapping of the atoms in the rf-dressed potential has been studied numerically with the help of a DSMC algorithm, after knowing the time dependent rf-dressed potential $V(\textbf{r},t)$ and LZ transition probability for the atom. An atom interacting with $V(\textbf{r},t)$, the evolution of the Bosonic ensemble distribution function $f(\textbf{p},\textbf{r},t)$ is given by the Boltzmann equation in terms of position (\textbf{r}), momentum (\textbf{p}) and time (t) as \cite{Wade:2011},
\begin{equation}\label{eq:BE}
\left[\frac{\partial}{\partial t}+\frac{\textbf{p}}{m}.\nabla_r-\frac{1}{m}\nabla_rV(\textbf{r},t).\nabla_p\right]f=I[\textit{f}\ ].
\end{equation}
The collision integral $I[f]$ can be calculated in terms of the differential scattering cross section $\frac{d\sigma}{d\Omega}$ and two distribution function $f_1$ and $f_2$ by using the equation,
\begin{equation}\label{eq:I}
I\left[\textit{f}\ \right]=\frac{1}{m}\int\frac{d^3p_2}{h^3}\int d\Omega\frac{d\sigma}{d\Omega}|\textbf{p}_2-\textbf{p}_1|[f_1'f_2'-f_1f_2],
\end{equation}
where m is the mass of the atoms, $\textbf{p}_1$ and $\textbf{p}_2$ are the momentum of the two atoms with their respective distribution functions $f_1$ and $f_2$. The primed notation denotes distribution functions after collision. The differential cross section can be calculated as,
\begin{equation}
\frac{d\sigma}{d\Omega}=|f_{sc}(\theta)\pm f_{sc}(\pi-\theta)|^2
\end{equation}
where the scattering function $f_{sc}(\theta)$ is given as,
\begin{equation}
f_{sc}(\theta)=\frac{\hbar}{imv_r}\sum_{l=0}^{\infty}(2l+1)(e^{2i\delta_l}-1)P_l(cos\theta),
\end{equation}
with $v_r$ is the relative velocity of colliding particles, $\delta_l$ is the phase shift associated with partial wave $l$, $P_l(cos\theta)$ is the $l^{th}$ order Legendre polynomial and $\theta$ is the center-of-mass scattering angle.

The alternative approach to solve above equations for continuous distribution function $f(p,r,t)$ is to replace this ensemble with a swarm of test particles having discrete positions and momentum consistent with the initial trap configuration and temperature, as per the Bird's proposal \cite{Bird:1994}. These test particles can represent more than one atoms in the ensemble as ``macro atoms" and reduce the simulation cost of large atom clouds significantly. DSMC model simulates the phase space variables, \textit{i.e.} position and momentum, of all atoms in the ensemble for all time steps during the evolution time. The key approximation in this method is the decoupling of collision free evolution of the atomic trajectories with the trajectories incorporating inter-atomic collisions. The dynamics has been studied by increasing the time in steps smaller than the mean collision time to ensure the validity of the decoupling approximation.

To begin with, the whole initial atom cloud is divided spatially over many cells to gain uniformity in the density and trapping potential over a cell. These cells are further split in sub-cells to increase the efficiency of the collision detection algorithm and parallel processing \cite{Wade:2011}. In order to get the collision-less trajectories of the atoms, a symplectic algorithm has been used which can conserve the total energy of the system over a time scale much longer than the total simulation time. To evaluate the phase-space variables in collision-less evolution over a time interval $\Delta t$, the following set of equations in terms of position (\textbf{r}), momentum (\textbf{p}) and time estimated position (\textbf{q}) are used for any $\textit{i}^{th}$ atom in the ensemble,  

\begin{equation}\label{eq:update}
\begin{aligned}
 & \textbf{q}_i=\textbf{r}_i(t)+\frac{\Delta t}{2m}\textbf{p}_i(t),\\
 & \textbf{p}_i(t+\Delta t)=\textbf{p}_i(t)-\Delta t\nabla_{q_i}V(\textbf{q}_i,t),\\
 & \textbf{r}_i(t+\Delta t)=\textbf{q}_i+\frac{\Delta t}{2m}\textbf{p}_i(t+\Delta t).
\end{aligned}
\end{equation}

After deriving the new phase space coordinates of all the atoms in the ensemble, the collisions are considered by calculating the collisional probability between the $\textit{i}^{th}$ and $\textit{j}^{th}$ atom in a cell as,
\begin{equation}
P_{ij}=\alpha\frac{\Delta t}{\Delta V_c}v_r\sigma(v_r),
\end{equation}
where $\alpha$ denotes the ratio between the number of test particles to the actual number of atoms simulated, $\Delta V_c$ is the volume of the cell and $\sigma(v_r)$ is the total cross section. The actual collision event is then performed based on an acceptance-rejection method. The momentum components of the colliding atoms are then updated with new momentum values calculated from the classical momentum conservation principle. Repeating this procedure over all the time steps constituting the total simulation time results in the complete dynamics of the atom cloud in the trapping potential.

Further, the DSMC method calculates all the positions of all the atoms, irrespective of whether an atom, after the simulation contributes to the formation of the cloud or not. Hence a region of interest is set to count the number of atoms in the vicinity of the main atom cloud. This region is taken wide enough to avoid any critical dependence of the atom number on the selected region. Hyperfine spin states are treated semi-classically with all the atoms initially in the  sate $|m_F=-2\rangle_A$ and any atom changes its state only when the transition probability $P_{LZ}$ of the corresponding atom passes an acceptance-rejection algorithm. 

We note that finding solution of Eq. (\ref{eq:update}), instead of solving Eq. (\ref{eq:BE}) which describes the continuous evolution of the distribution $f(\textbf{p},\textbf{r},t)$, can result in fluctuations other than the physical ones. To minimise these fluctuations, simulations are repeated for a number of times and average results are used for the study.

\section{Results and discussion}\label{Results}

\subsection{Experimental procedure and observations}

\begin{figure}[b]
\includegraphics[width=8.6cm]{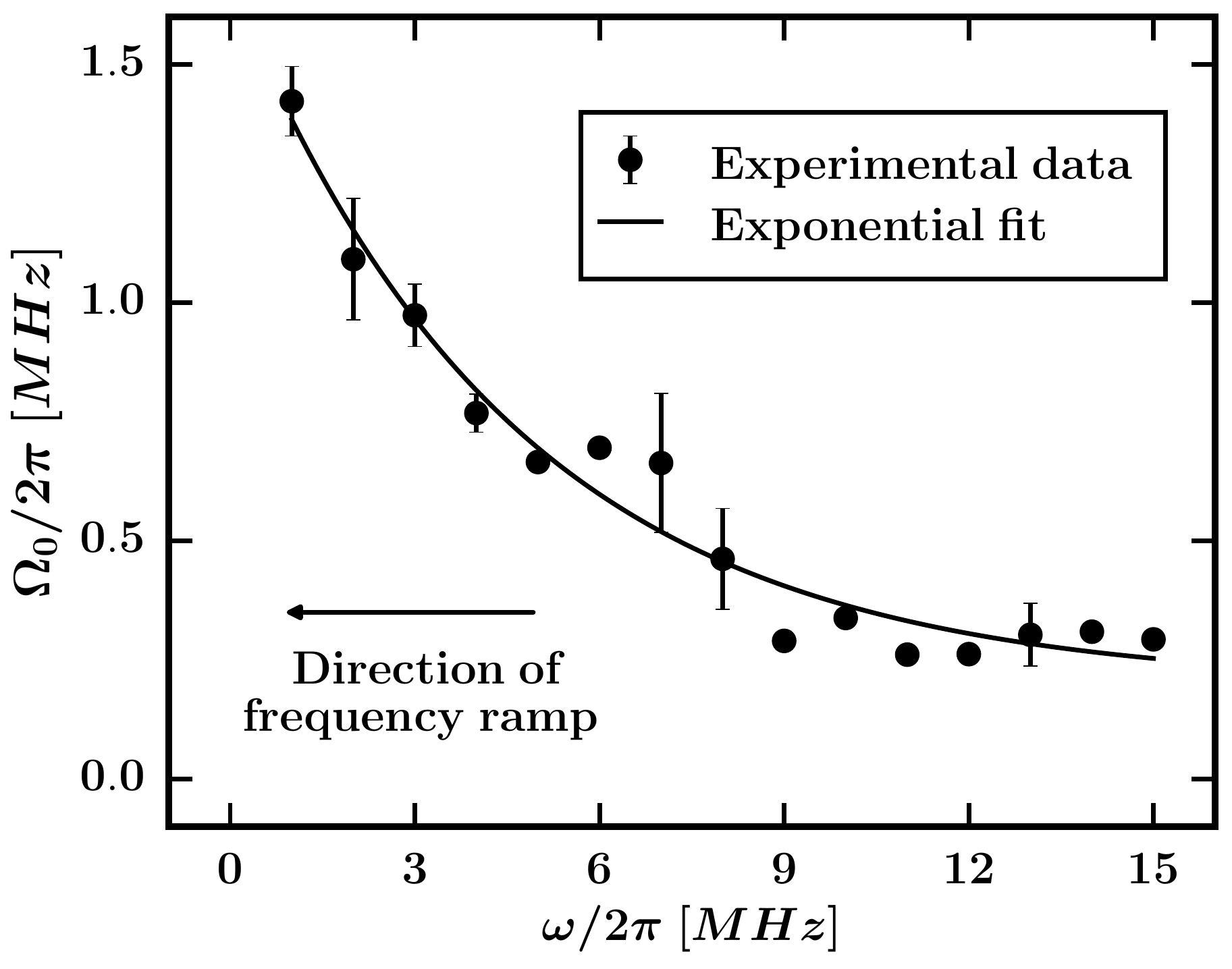}
\caption{\label{fig:Omega} Measured variation in the peak Rabi frequency $\left(\Omega_0=(g_F\mu_B/\hbar)|B^{rf}|\right)$ as a function of rf-field frequency during the frequency ramp. The continuous curve is an exponential fit. The error-bar shows the standard deviation between multiple experimental observations.}
\end{figure}

\begin{figure}[t]
\includegraphics[width=8.6cm]{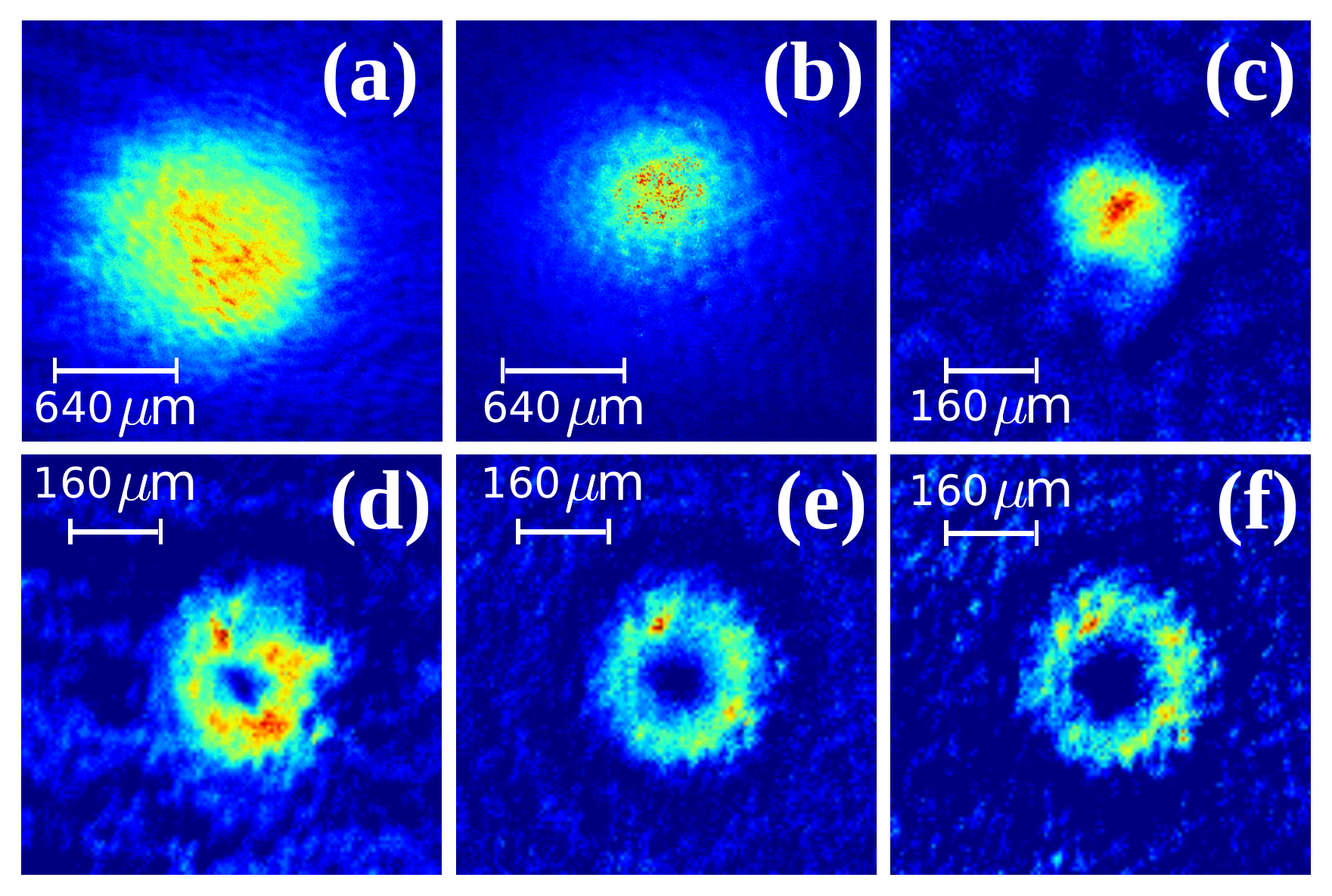}
\caption{\label{fig:5sall}(Color online) Experimentally observed absorption images of the cold atom cloud at different instants of time after starting the ramp of the rf-field : (a) t = 1 sec, (b) t = 1.5 sec, (c) t = 2.5 sec, (d) t = 3.0 sec, (e) t = 4.0 sec and (f) t = 5.0 sec. The colours from blue to red in all the images show optical density in increasing order.}
\end{figure}

The experiments have been performed on a double magneto-optical trap (double-MOT) setup which is described in detail elsewhere \cite{Chakraborty:2016}. In this setup, a vapor chamber MOT (VC-MOT) is prepared in an octagonal chamber of stainless steal (SS) which has Rb-vapor in the background. The second MOT, called UHV-MOT, is prepared in a glass cell at a pressure of $\sim 5\times 10^{-11}$ Torr. The UHV-MOT is loaded by transferring atoms from the VC-MOT using a push laser beam. Nearly $2\times 10^8$ atoms at temperature $\sim$250 $\mu K$ are collected in the UHV-MOT. The UHV-MOT atoms are then trapped in a quadrupole magnetic trap, after following a sequence of processes such as compressed-MOT, optical molasses and optical pumping to $|F =2, m_F =2 \rangle$. Nearly $2\times 10^7$ atoms at temperature  $\sim$ 200 $\mu K$ were trapped in the quadrupole trap after appropriately ramping-up current in the trap coils (to achieve final axial field gradient of 180 G/cm). After the complete formation of the quadrupole trap (which includes 50 ms time for settling the atom cloud in the trap), an rf-field of variable frequency $(\omega/2\pi)$ and amplitude $B^{rf}$ is applied on trapped atom cloud using a 10-turn loop antenna (2 cm diameter) with its axis along the x-axis. The rf-field is ramped (varied with time) in a chosen way to convert quadrupole trap into an rf-dressed toroidal trap. The axis of the quadrupole trap is along the z-axis and the gravity is along the direction opposite to the y-axis. By applying a resonant probe beam along the z-axis, the absorption images of the trapped atom cloud were recorded. As the absorption probe imaging is an invasive technique, a fresh cycle of experiment starting from MOT loading to rf-field exposure is run to capture the image of the atom cloud after changing any experimental parameter. In each experimental cycle, the image is recorded by setting an appropriate delay in CCD trigger with respect to start of the frequency ramp of rf-field.

During the ramp of the rf-field in our experiments, the frequency $(\omega/2\pi)$ of rf synthesizer was varied linearly with time from 15 MHz to 1 MHz in 5 sec duration. Using an appropriately designed impedance matching circuit for the rf-amplifier, the rf-field amplitude $B^{rf}$ was increased exponentially from 0.1 G to 2 G during the above frequency variation in the field. Fig. \ref{fig:Omega} shows the variation in the rf-field strength (in terms of peak Rabi frequency $\Omega_0=(g_F\mu_B/\hbar)|B^{rf}|$) as function of the rf-field frequency $(\omega/2\pi)$ in the ramp. Applying this ramp of the rf-field, the atom cloud trapped in the quadrupole trap was transferred to a toroidal/ring trap in our experiments. In the measurements, this rf-field strength $B^{rf}$ was estimated in the long wavelength approximation by measuring the rf-current passing through the loop antenna. We note that such a high value of $B^{rf}$ corresponds to the case of non-RWA regime when atom field interaction is considered. The high value of Rabi frequency enhances the non adiabatic Landau Zener tunnelling rate that transfers atoms to trappable dressed state $|m_F=2\rangle_A$. During the rf ramp, the temperature of the atom cloud reduces to $\sim$ 40 $\mu K$ in the rf-dressed trap, after the evaporative cooling phase, from the initial $\sim$ 200 $\mu K$ temperature in the quadrupole trap. As shown in Fig. \ref{fig:5sall} (images (d)-(f)), the radius of ring increases with time during the ramp. This can be understood as follows. In the higher frequency regime, \textit{i.e.} in the beginning of the ramp, the lower coupling strength limits the height of the central hump in the rf-dressed potential for the dressed state $|m_F=2\rangle_A$. This low height of the potential hump is not sufficient to push atoms away from the centre of the quadrupole trap. As frequency decreases further during the ramp, the coupling strength increases, which gives rise to an increase in the height of the central hump in the potential. As this height increases, the cold atoms are pushed away from the centre, which leads to increase in the radius of the ring cloud.

\begin{figure}[t]
\includegraphics[width=8.6cm]{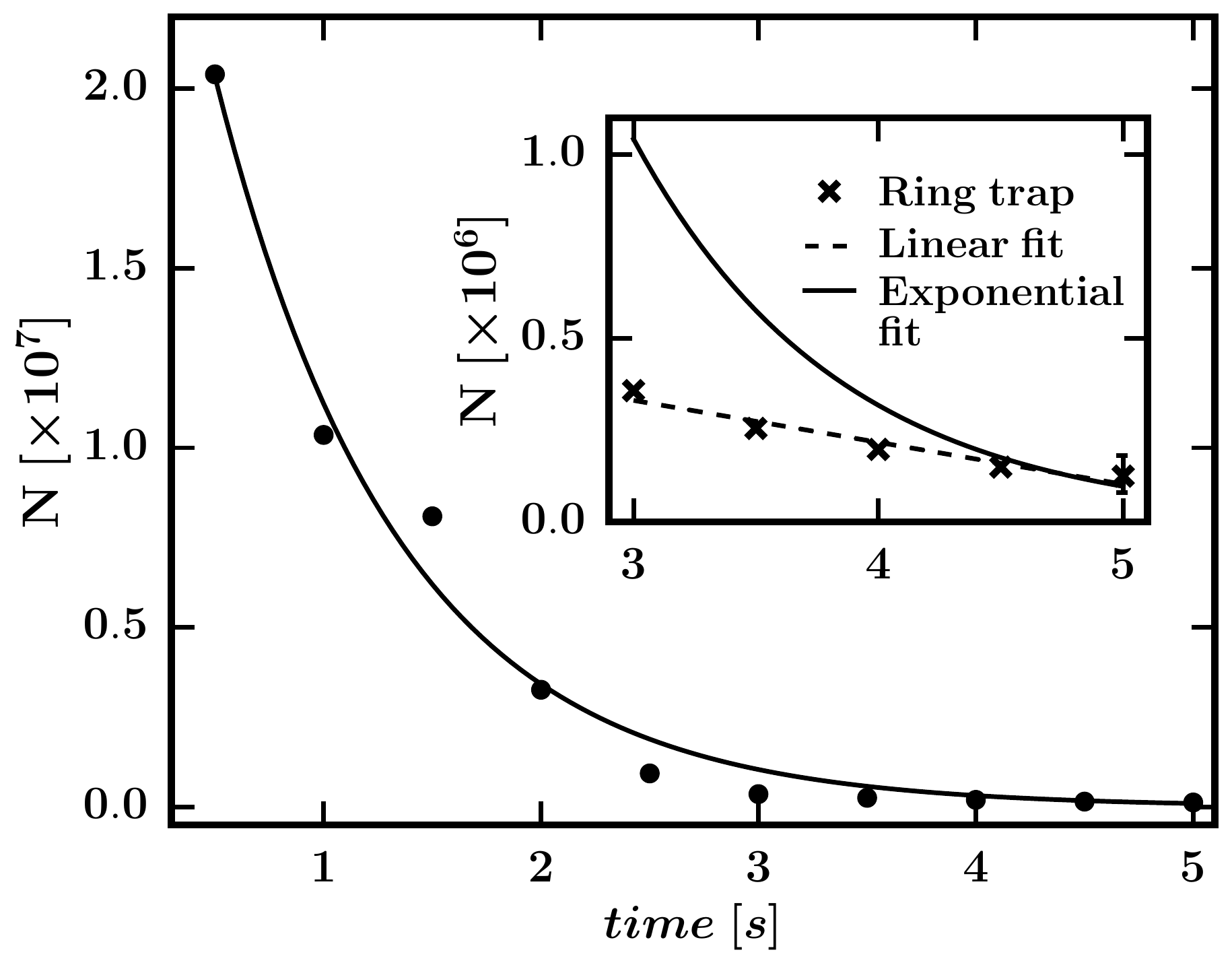}
\caption{\label{fig:5sN} The measured variation in number of atoms (dark circles) in a trapped cloud during the ramp down of frequency from 15 MHz to 1 MHz in 5 sec duration. The initial faster decay of number of atoms in the trap bears the signature of the rf induced forced evaporation, whereas the slow variation at the trailing edge shows the trapping of atoms in the rf-dressed potential having lower losses. The continuous curve shows the exponential decay in number of atoms. The inset shows the magnified view of the evolution of number of atoms in the rf-dressed ring potential.}
\end{figure}

In experiments the atom cloud shows a fast decrease in number of trapped atoms with time, in the beginning of the rf ramp. This bears the signature of forced evaporative cooling of the atom cloud. However, in the later part of the ramp, as the frequency of the rf-radiation reaches to $\sim$ 6 MHz at time $\sim$3 sec, the exponentially increasing coupling strength dominates to modify the potential landscape. The high value of coupling strength results in trapping of atoms in the rf-dressed potential in the toroidal geometry. This is the cause of the reduced loss rate of atoms after 3 sec of frequency ramp, as shown in Fig. \ref{fig:5sN}. This is also evident from the Fig. \ref{fig:5sall} (images (d)-(f)) which shows the reshaping of atom cloud with low density in the central regime, indicative of trapping in the dressed state potential after duration of 3 sec.

\subsection{Numerical results and comparisons}

\begin{figure}[b]
\includegraphics[width=8.6cm]{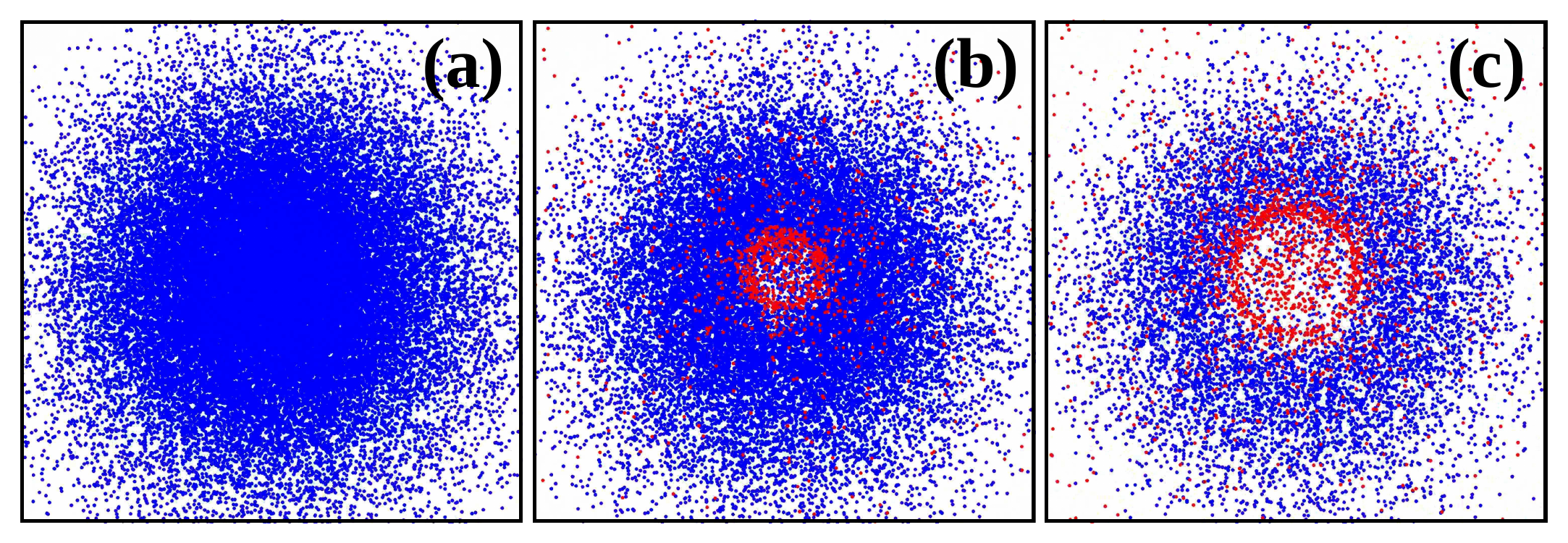}
\caption{\label{fig:simulation}(Color online) Simulated atom cloud pictures at different instants of time during the rf-frequency ramp from 15 MHz to 1 MHz in 5 sec. The simulation results are for initial number of atoms $5\times 10^4$ at temperature 200 $\mu K$ in quadrupole trap. The blue colour dots represent particles in the $|m_F=-2\rangle_A$ state and red dots represent particles in $|m_F=2\rangle_A$ state. The ring with red dots indicates the cloud trapped in the toroidal geometry. Figures (a) to (c) show the simulated cloud shapes as frequency ramp progresses.}
\end{figure}

\begin{figure}[t]
\includegraphics[width=8.6cm]{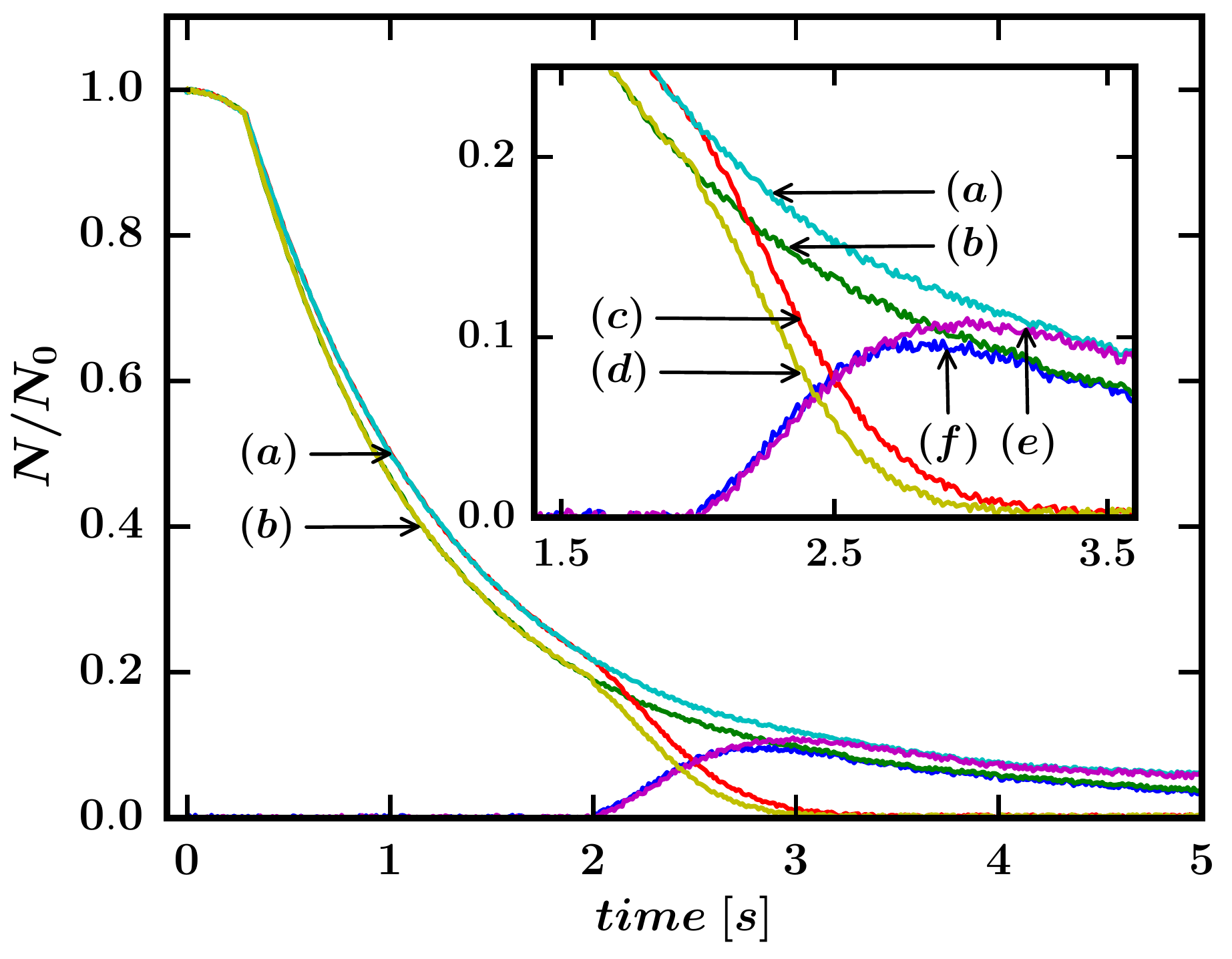}
\caption{\label{fig:mdsimN}(Color online) Results of numerical simulations showing the variation in number of atoms with time during the ramp down of rf-field frequency. The curves (a), (c) and (e) show results of simulations for initial number $N_0$=$1\times 10^6$, whereas (b), (d) and (f) show the results for $1\times 10^5$. The curves (c) and (d) show number of atoms in $|m_F=-2\rangle_A$ state whereas (e) and (f) show atom numbers in $|m_F=2\rangle_A$ state. The inset shows the magnified view of the evolution of number of atoms in different states.}
\end{figure}

Numerical simulations have been performed to understand these experimental results by considering the interaction of the trapped atom cloud with an rf-field of time varying amplitude and frequency. Similar to our experimental conditions, the frequency of the rf-field was ramped down linearly with time and the corresponding coupling strength was varied exponentially as a function of the rf-frequency and shown by the continuous curve in Fig. \ref{fig:Omega}. We start from an atom cloud with randomized Gaussian distribution both in position and momentum space, where the width parameters ($\sigma_i$,$\sigma_{v_i}$, $i=x,y,z$) are governed by the temperature of the cloud (200$\mu K$) and the magnetic field gradient of the trap. All the parameters used in the simulations were chosen close to their experimental values, except the number of atoms. The number of atoms was lower than the experimental value due to the limited computational speed. At each time interval, the total number of trapped atoms were calculated by the region of interest method, described in the earlier section along with their state. To start with, all the atoms were considered in $|m_F=-2\rangle_A$ state, which is confined by a rf-dressed state potential having a single minimum at the quadrupole trap centre as shown in Fig. \ref{fig:simulation} (a). The atom cloud in this potential has a nearly Gaussian profile for the number density, similar to that in the bare quadrupole trap. As the frequency of rf-field is ramped, the parameters of rf-field responsible for trapping atoms in the state $|m_F=-2\rangle_A$ change as shown in Fig. \ref{fig:Omega}. Initially at high frequency and low amplitude of rf-field, the evaporation process takes place which removes atoms having velocity higher than the average velocity. As the frequency decreases and the amplitude increases further, the non-adiabatic LZ transition probability to $|m_F=2\rangle_A$ state is increased. This results in trapping atoms in a different potential landscape (Fig. \ref{fig:field_line}) corresponding to this new state $|m_F=2\rangle_A$, which is reflected in the number density and cloud shape. Fig. \ref{fig:simulation} (b) and (c) shows the evolution in the shape and density distribution of the simulated atom cloud.   

The variation in the number of atoms trapped during the frequency ramp process is shown in Fig. \ref{fig:mdsimN}. An important feature here is the change in the loss rate during the frequency ramp, with a higher loss rate in the beginning of the ramp than that towards the end of the ramp (Fig. \ref{fig:mdsimN}) similar to the experimental observations depicted in Fig. \ref{fig:5sN}. This is indicative of trapping atoms in rf-dressed potential towards the end of the ramp, after initial loss due to rf-field induced evaporation. The rise of population in $|m_F=2\rangle_A$ state in Fig. \ref{fig:mdsimN} after 3 sec supports this argument. Towards the end of the frequency ramp process, the population in $|m_F=-2\rangle_A$ state drops to an extremely small fraction of the initial total number of trapped atoms, and population in $|m_F=2\rangle_A$ state reaches the maximum value and becomes the sole contributor to the trapped atom cloud. The hollowness of the atom cloud in Fig. \ref{fig:simulation} suggests the trapping of atoms in the potential landscape of the dressed state $|m_F=2\rangle_A$.

\begin{figure}[t]
\includegraphics[width=8.6cm]{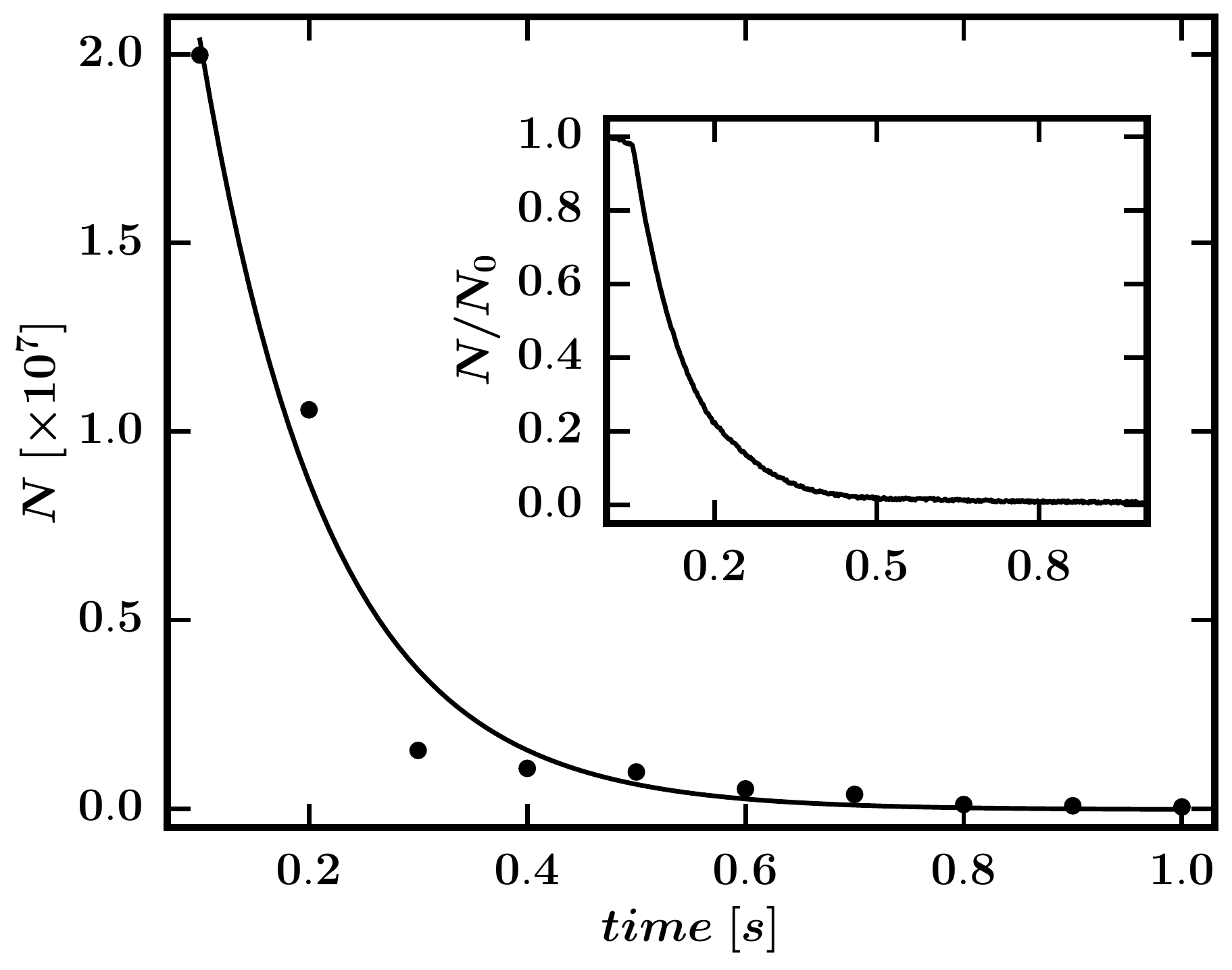}
\caption{\label{fig:1sN}(Color online) The measured variation in number of atoms (circles) in the trapped cloud with time when frequency of rf-field is ramped down from 15 MHz to 1 MHz in 1 sec. The continuous curve is exponential fit to the data and guide to eyes. The inset shows a similar variation obtained from simulations for $N_0=1\times 10^5$ atoms.}
\end{figure}

\subsection{Remarks}

In order to explore the results with a more fast ramp down of frequency of the rf-field, a similar experiment has been conducted where the frequency ramp down from 15 MHz to 1 MHz was achieved in a 1 sec duration. This fast ramping resulted a very fast decrease in the number of atoms in the trap (as shown in Fig. \ref{fig:1sN}), with final number of atoms remaining in the trap after 1 sec was only $0.2\%$ of the initial number. At the same time, the formation of the ring trap was also not observable. The numerical simulations for this fast ramp down (\textit{i.e.} 1 sec) have also provided the similar results, which shows the ability of our simulation to handle such a fast and non-equilibrium trap dynamics.

Finally, we wish to remark that a further lower temperature in rf-dressed ring trap potential can be achieved by applying a weak rf field to induce transitions between the dressed states. The rf-field for that can have a frequency close to the separation between the dressed energy levels  \citep{Easwaran:2010}. Due to high Rabi transition rates in the obtained dressed trap, this weak rf-field under the non-RWA regime excites population to multiple dressed manifolds and results in a decrease of the $|m_F=2\rangle_A$ population at frequencies $n\omega+\Omega$ ($n$=0,1,2,..). Such excitations are not usually accessible for the trapping in RWA regime. Presence of these resonances makes this trap a wonderful tool, not only to study higher order dressed state transitions, but also to incorporate more intricate and effective evaporation mechanisms hitherto not studied. 

\section{Conclusion}\label{Conclusion}
Using the Direct Simulation Monte Carlo (DSMC) technique, we have studied the dynamics of an atom cloud during the conversion of its trapping potential from a quadrupole magnetic trap to an rf-dressed potential of toroidal shape. This rf-dressed potential for the toroidal trapping of atoms is formed when atoms trapped in a quadrupole trap are exposed to an rf-field with time varying frequency and amplitude. The results of simulations have provided an insight into the processes governing the atom cloud parameters. It is found that, initially at low rf coupling strength, the evaporative cooling dominates, whereas at higher coupling strength trapping of atoms occurs in the rf-dressed potential. The predictions of the simulations have been found adequate to explain the experimental observations. The work presented in this article can be extended to study the non-equilibrium trap dynamics with more sophisticated evolution of trapping potential, including the potential evolution beyond the rotating wave approximation. Using similar simulations more useful and effective evaporation and rf-dressing schemes can be explored.

\section*{Acknowledgement}
We thank A. Srivastava for her help during the experiments. We also thank M. Lad and P. S. Bagduwal for providing the rf amplifier and A. K. Pathak, S. Tiwari and L. Jain for modifications in the control system for the experiments. A. Chakraborty acknowledges the financial support by the Homi Bhabha National Institute, India.

\appendix*
\setcounter{equation}{0} 
\section*{Appendix}

The interaction Hamiltonian can be written in terms of the detuning element $a=(|B^S(\textbf{r})|-\frac{\hbar\omega}{g_F\mu_B})$ and conjugate Rabi elements, $b=(B^{rf}_{\bot1}-iB^{rf}_{\bot2}e^{i\gamma})$ and $b^*$, 
\begin{equation}
H_I=\frac{H}{g_F\mu_B}=\left[\begin{array}{ccccc}
2a           & \frac{1}{2}b          & 0                       & 0                     & 0 \\ 
\frac{1}{2}b^* & a                     & \frac{\sqrt{6}}{4}b     & 0                     & 0 \\ 
0            & \frac{\sqrt{6}}{4}b^*   & 0                       & \frac{\sqrt{6}}{4}b   & 0 \\ 
0            & 0                     & \frac{\sqrt{6}}{4}b^*     & -a                    & \frac{1}{2}b \\ 
0            & 0                     & 0                       & \frac{1}{2}b^*          & -2a
\end{array}\right]. 
\end{equation}

Diagonalizing this Hamiltonian one can obtain all the associated dressed eigen values ($\lambda$),
\begin{equation}
\lambda=0,\pm1\sqrt{a^2+\frac{|b|^2}{4}},\pm2\sqrt{a^2+\frac{|b|^2}{4}},
\end{equation}
and eigen states in a compact form,
\begin{equation}\label{eq:dressedstates}
|m_F=j\rangle_A = \sum_{k=-F}^{F}C_k^j|m_F=k\rangle,\ for\ j=-F,...,F
\end{equation}
where the subscript A stands for the adiabatic dressed states. The coefficients $C_k^j$ are the matrix element of an unitary matrix provides mapping between the bare and dressed state basis. All the $F\times F$ $C_k^j$ coefficients can be written in terms of a, b and $d=\sqrt{4a^2+|b|^2}$. For a particular energy eigen value ($\lambda=0$) the calculated coefficients are,

\begin{align}
	& C_2^0=\frac{\sqrt{3}|b|^2}{D}, C_1^0=-\frac{4\sqrt{3}ab}{D}, \nonumber \\
	& C_0^0=\frac{\sqrt{2}(8a^2-|b|^2)}{D},C_{-1}^0=\frac{4\sqrt{3}ab^*}{D} \nonumber \\
	& C_{-2}^0=\frac{\sqrt{3}|b|^2}{D}, D=[96a^2b^2+2(8a^2-b^2)^2+6b^4]^{\frac{1}{2}} \nonumber \\ 
\end{align}

All other coefficients corresponding to other eigen values can also be evaluated by following the same procedure. As these coefficients entail the contribution of any bare state in different dressed states, determination of a coupling strength suitable for either the evaporation or the dressed state requires the determination of the coefficients also.

%\bibliographystyle{apsrev}
%\bibliography{Reference}

\hspace{0.1cm}

\end{document}